\documentclass[%
reprint,%
amssymb, amsmath,%
aip,cha,%
]{revtex4-1}
\usepackage[T1]{fontenc}
\usepackage[latin9]{inputenc}
\usepackage{color}
\usepackage{units}
\usepackage{amssymb}
\usepackage{graphicx}
\usepackage{esint}
\usepackage{bm}
\usepackage{natbib}
\usepackage{ulem}
\usepackage[colorlinks=true, citecolor=red]{hyperref}
\setcitestyle{journalcolor= blue}


\begin{document}
		
	\title { Anomalous photoluminescence emission of monolayer MoS$_2$-QD heterostructure on hBN.}

	\author{H L Pradeepa}
	\email{pradeepa@alum.iisc.ac.in}
	\affiliation{Department of Physics, Indian Institute of Science, Bangalore 560012, India}

\begin{abstract}
	
Monolayer transition metal dichalcogenides(2D) and zero dimensional quantum dots(QD) are known to have unique optical properties in their individual limit such as high binding energy of excitons. The combination of these two systems is of particular interest in understanding various aspects of energy transfer, charge transfer, dynamics of excitons, etc. In this manuscript, we report the  anomalous photoluminescence(PL) emission   in one such heterostructure MoS$_2$-CdSe QD. We observe multiple exciton emission peaks of the heterostructure   on hBN substrate which are absent on SiO$_2$. Our observation open up the questions, whether the local potential due to the lattice mismatch between MoS$_2$ and hBN has any role in deciding the  emission of these peaks  or the strain field of MoS$_2$ and hBN is the reason for the emergence of multiple emission. In addition,    the altered quantum potential of QD due the presence of hBN and MoS$_2$ may also leads to such multiple emissions. 
  
\end{abstract}
			
\maketitle


The recent studies of two-dimensional semiconductors(2D) and their hybrid structure with zero-dimensional(0D) semiconductors have led to discovering of many fascinating properties that are absent in their bulk counterparts.\cite{wang2018colloquium,ross2013electrical,bera2010quantum,raja2016energy} MoS$_2$ monolayer, one such 2D semiconductor  is known to have a very interesting properties such as transition from indirect to direct band gap from bulk(1.2 eV) to monolayer(1.8 eV) limit, high binding energy of excitons(0.4 to 0.6 eV) and trions(30 to 40 meV) at room temperature.\cite{ramasubramaniam2012large} 
We can easily tune the PL emission of CdSe	QDs from 2.4 eV to 1.8 eV,\cite{pradeepa2020strong,haridas2013photoluminescence} where as the absorption and PL emission of monolayer MoS$_2$	lies in between 2.15 eV to 1.7 eV,\cite{mak2013tightly} this makes the CdSe QDs a suitable 0D emitter with monolayer MoS$_2$ to make 2D-0D hetero-structure and study their exotic properties.\cite{alexeev2019resonantly,jin2019observation,tran2019evidence,seyler2019signatures}
From the fundamental perspective, interest has been focused on the novel aspects of light-matter interactions that can occur between the two nanoscale materials in the form of energy and charge transfer processes between photo excited excitons which can be generated in one or both the layers while under certain conditions formation of interlayer or hybrid excitons can also take place.\cite{boulesbaa2016ultrafast,alexeev2019resonantly}   

 Here we report the low temperature PL study of  monolayer MoS$_2$(2D)- CdSe (0D) hybrid structures. We observed multiple exciton emission peaks in  the heterostructure of MoS$_2$-QD in which hBN is used as underneath substrate. These emissions   on hBN  are absent on SiO$_2$ substrate region. Our observation opens the questions whether the local potential due to the lattice mismatch between MoS$_2$ and hBN has any role in deciding the  emission of these peaks  or the strain field of MoS$_2$ and hBN is the reason for the emergence of multiple emission. The altered quantum potential of QD due the presence of hBN and MoS$_2$ may also contribute to these multiple emission mechanism.

\begin{figure}[t]
	\includegraphics[width=1.0\linewidth]{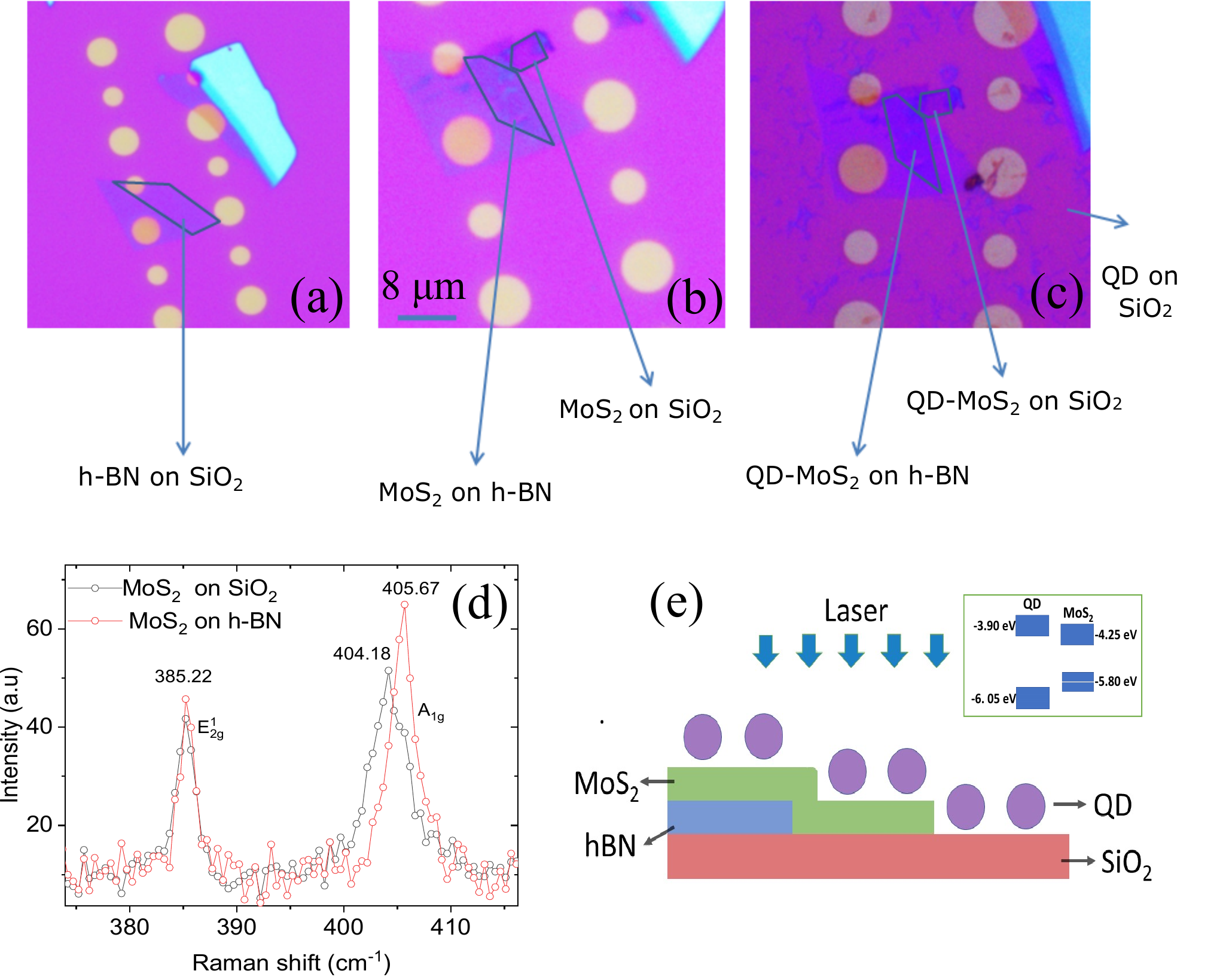}
	\caption{ \label{fig:sc} (a) Optical image of h-BN on SiO$_2$.(b) Optical image of MoS$_2$ on h-BN and on SiO$_2$.(c) Optical image of QD-MoS$_2$ heterostructure on h-BN and on SiO$_2$. (d) Room temperature Raman spectra of the monolayer MoS$_2$ on SiO$_2$ and on h-BN. (e) Schematic of the experimental set-up. Inset shows the band diagram of QD and MoS$_2$, the energies of valance and conduction band of both QD and MoS$_2$ favor the possibility of interlayer excitons. }
\end{figure} 

h-BN flakes were exfoliated on polydimethylsiloxane (PDMS) sheets and then transferred on  300 nm SiO$_2$ substrates,       fig.~\ref{fig:sc}(a) shows the optical image of h-BN transferred on SiO$_2$.  MoS$_2$ monolayer flakes were prepared   using standard exfoliation technique  on  PDMS sheets. Optical microscopy and Raman spectroscopy were used to identify the monolayer.    MoS$_2$ monolayers were then  transferred  in such a way that  some portion of  MoS$_2$ is on h-BN and some portion on SiO$_2$ as shown  in Fig.~\ref{fig:sc}(b). CdSe QDs were  synthesized following methods described earlier  \cite{de2003single,qu2002control}. The QD monolayer  was transferred on MoS$_2$ using  Langmuir-Blodgett(LB) technique\cite{collier1997reversible,dabbousi1994langmuir,heath1997pressure}, using LB trough(Kibron Microtrough G2,Finland). Fig.~\ref{fig:sc}(c) shows the optical image of the QD-MoS$_2$ heterostructure on h-BN and on SiO$_2$.
 Fig.~\ref{fig:sc}(d) shows the Raman spectra of MoS$_2$ on SiO$_2$ and on h-BN at room temperature. In schematic fig.~\ref{fig:sc} (d) the vertical cross sectional view of the hetero-structure has been shown. As seen in the band diagram of QD and MoS$_2$, the energies of valance and conduction band of both QD and MoS$_2$ favor the possibility of interlayer excitons.

  PL and Raman spectra were collected   using the Horiba (LabRam model) instrument, using 532 nm continues wave (CW) laser   to excite the sample keeping the laser power  $\sim $ 2 $ \mu $W. Signals were collected  using charge coupled device(CCD). 300 g/mm grating and   1800 g/mm grating were used to collect the PL and the Raman spectra respectively. 50x (Olympus NA-0.45) objective was used  collect both PL  and  Raman data.  Montana (Cryostation model) was used mounted to Horiba system to collect  low temperature spectra.

    \begin{figure}[t]
    	\includegraphics[width=1\linewidth]{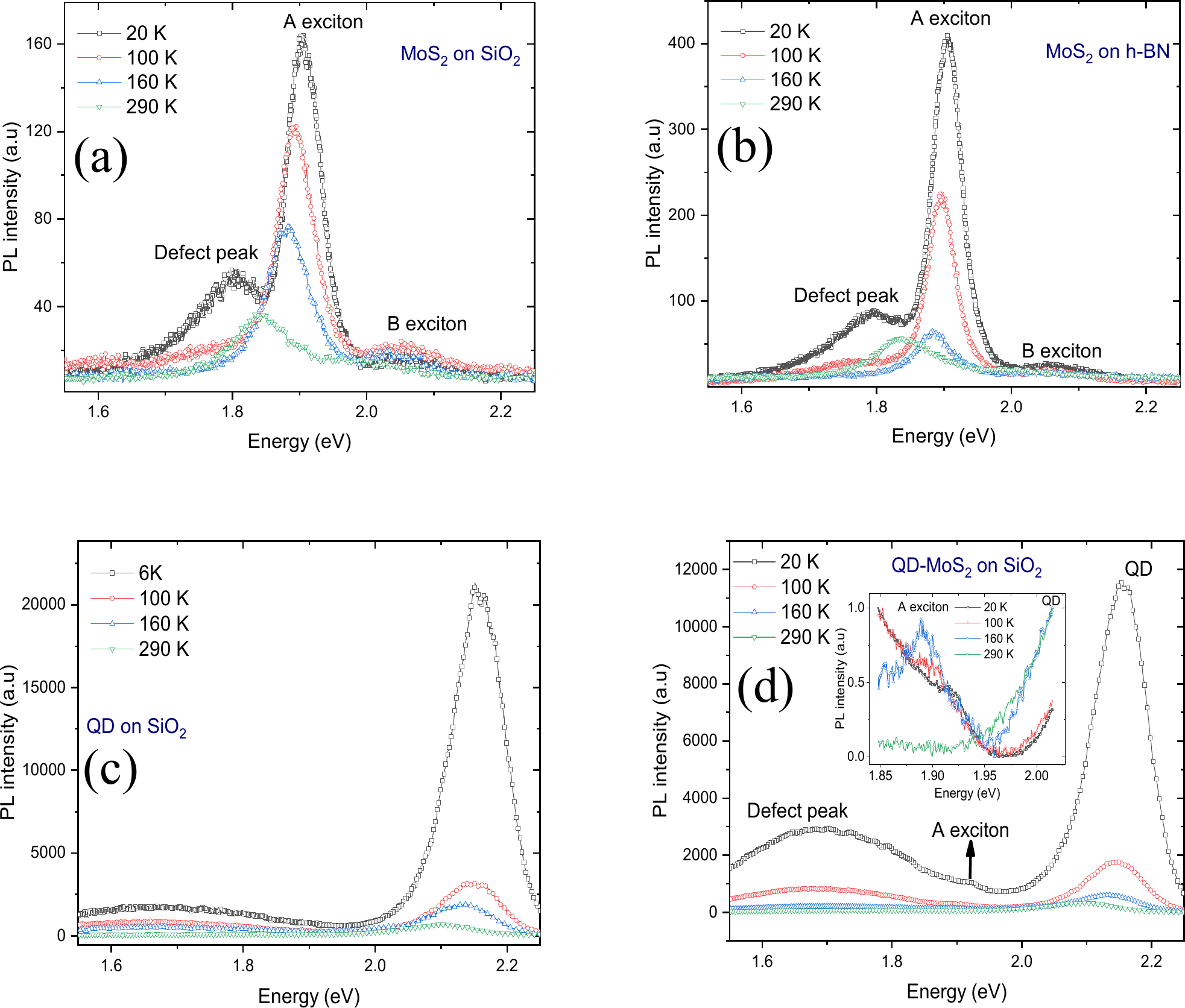}
    	\caption{ \label{fig:2}   Temperature dependent PL spectra: (a) MoS$_2$ on SiO$_2$, defect peak is observed at low temperatures at lower energy. (b) MoS$_2$ on h-BN, PL intensity of MoS$_2$ on h-BN was	increased compared MoS$_2$ on SiO$_2$. (c) QD on SiO$_2$, a broad defect peak is observed in QD PL spectra also at low temperatures. (d) QD-MoS$_2$ on SiO$_2$, inset shows the zoomed spectra	of MoS$_2$, where the B exciton peak is overlapped with the QD spectra.}
    \end{figure} 
    
    The PL emission spectra of monolayer MoS$_2$ at  K (K$'$) point consists of two peaks because of presence of  the strong spin orbital interaction  at around 1.88 eV (called A exciton) and 2.0 eV (called B exciton).  Fig.~\ref{fig:2}(a) shows the temperature dependent PL spectra of MoS$_2$ on SiO$_2$, defect peak is observed at low temperatures at lower energy. The A exciton PL is more sensitive compare to the B exciton PL. The PL intensity of A exciton increases with decreasing temperature(T). As   we decrease the T, we observe that the total PL intensity of A exciton increases  with   blue shift, further, this A exciton peak can be deconvoluted into exciton and trion peaks, the intensity of exciton  increases with decrease in T. This suggests that as we decrease the T the A exciton peak is more exciton in nature. The defect induced peak starts appearing at lower temperatures, this is due to the less available energy for the carriers at low T to overcome the trapping potential.   
     
     Fig.~\ref{fig:2}(b) shows the the temperature dependent PL spectra of MoS$_2$ on h-BN, it is observed that the PL intensity of MoS$_2$ on h-BN was increased compared to MoS$_2$ on SiO$_2$, this increase may be  due to substrate induced doping. Fig.~\ref{fig:2}(c) shows the  temperature dependent PL spectra of QD on SiO$_2$. The PL of QD increases with decrease in T, also the energy blue shifts as we decrease the T. We observe a broad defect peak  at lower energy in the PL at low temperatures which is associated to the trapping states.

     \begin{figure}[t]
     	\includegraphics[width=1\linewidth]{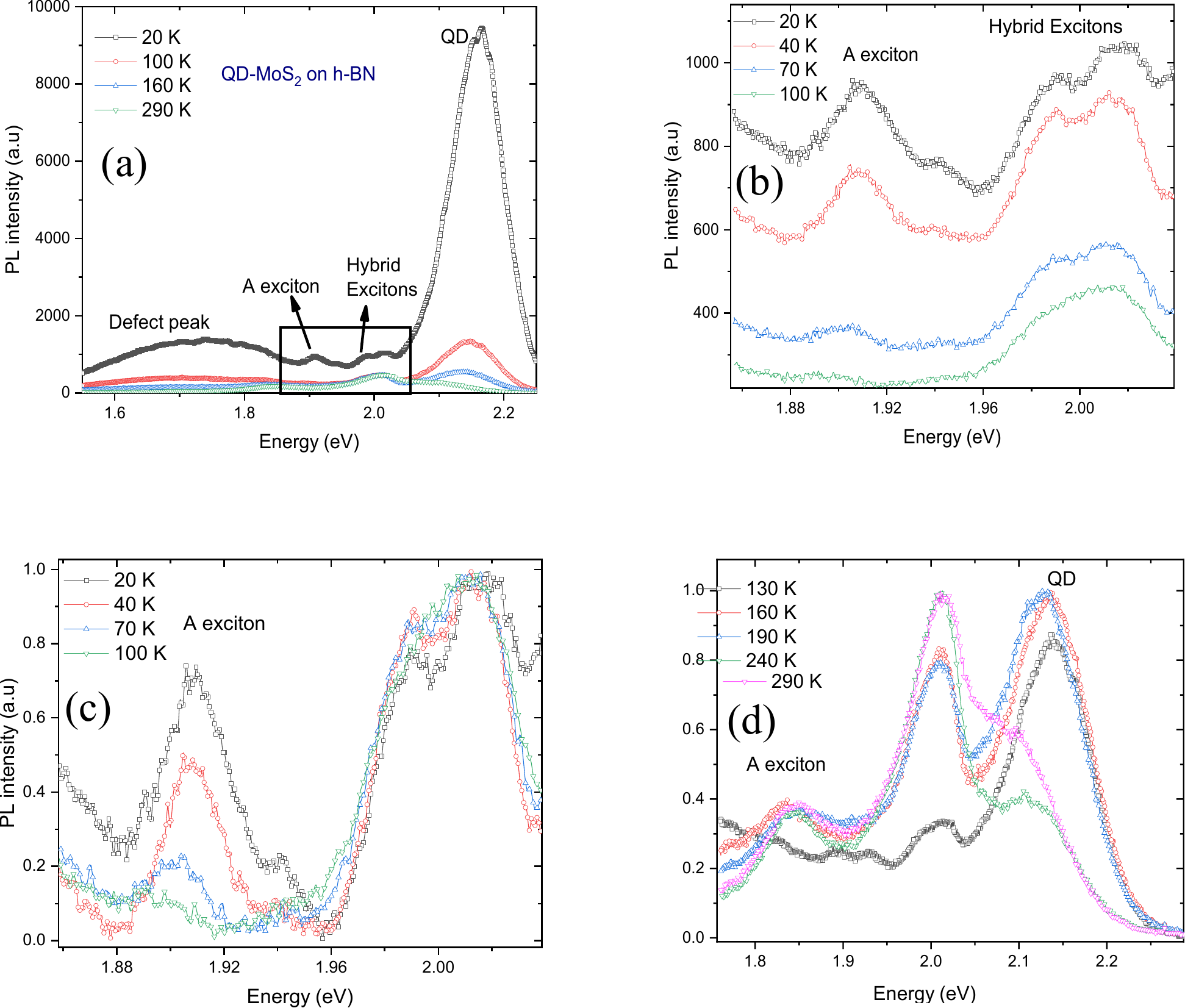}
     	\caption{ \label{fig:3} (a)  Temperature dependent PL spectra of QD-MoS$_2$ on h-BN. (a) QD-MoS$_2$ on h-BN	in the broad energy range. (b) PL spectra showing the zoomed region of MoS$_2$ emission from 20 K to 100 K. (c) Normalized spectra of zoomed regime K. (d) Normalized PL spectra	showing the of QD-MoS$_2$ from 130 K to 290 K.}  
     \end{figure}

       Fig.~\ref{fig:2}(d) shows the PL spectra of QD-MoS$_2$ on SiO$_2$ at different temperatures. MoS$_2$
       B exciton spectra is overlapped with the QD PL spectra, where as A exciton peak is still
       high enough to observe. QD peak is blue shifted and PL intensity is increased
       as we decrease the temperature. QD PL intensity decreases on MoS$_2$ indicating the energy
       transfer from QD. fig.~\ref{fig:2}(d) inset shows the zoomed spectra of MoS$_2$ in the QD-MoS$_2$ heterostructure spectra.

        Fig.~\ref{fig:3}(a) shows the temperature dependent PL spectra of QD-MoS$_2$ heterostructure on h-BN. Very interestingly, along with the blue shift and increase in PL intensity we observe extra peaks near the
        A and B exciton peaks of MoS$_2$. 
        Fig.~\ref{fig:3}(b) shows the temperature dependent PL spectra of QD-MoS$_2$ heterostructure on h-BN zoomed near the A and B excitons of MoS$_2$ from 20 K to 100 K. Between A exciton(energy-1.90
        eV) and QD (energy-2.16 eV) multiple peaks are observed at 1.94 ev, 1.99 ev and 2.02 eV.  As we increase the temperature
        the A exciton peak decreases, also the intensity of the higher energy multiple exciton peaks     decrease .

         For clarity, we plotted the  
         normalized PL spectra of QD-MoS$_2$ heterostructure on h-BN.  
         We can clearly see that the ratio of A exciton to multiple excitons peaks increases with increasing the temperature till
         100 K as shown in the normalized spectra in fig.~\ref{fig:3}(c). The peak near 1.99 eV starts merging
         with the peak near 2.02 eV after 100 K, where as the peak near 1.94 eV disappear after 130 K. Normalized PL spectra from 130 K to 290 K as shown in Fig.~\ref{fig:3}(d). The intensity of the higher energy  peak at 2.02 eV increases with
        increasing the temperature from 130 K to 290 K. Interestingly this  exciton peak  
        dominates the QD spectra at higher temperatures. More interestingly this peak
        is blue shifted with increasing the temperature.

\begin{figure}[t]
    	\includegraphics[width=0.9\linewidth]{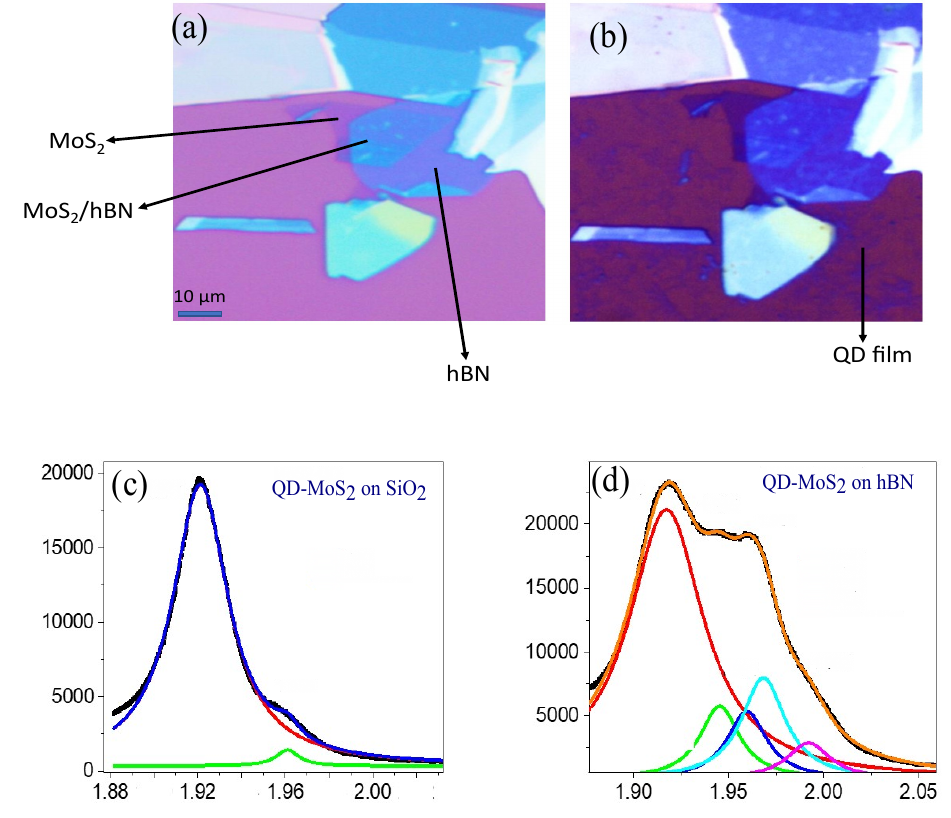}
    	\caption{ \label{fig:4}   Optical images of the second sample   before (a) and after (b) transferring the QD. (c) shows the PL spectra of QD-MoS$_2$ on SiO$_2$. (d) PL spectra QD-MoS$_2$ on hBN, multiple PL emissions  are resolved by fitting the PL spectra using multiple Lorentzian.  }
    \end{figure}

       In another sample we observed similar multiple PL peaks at low T at higher powers which   is shown in Fig.~\ref{fig:4}. Fig.~\ref{fig:4}(a) and (b) show the optical images of the  MoS$_2$ on hBN heterostructure before and after transferring QDs respectively. Fig.~\ref{fig:4}(c) shows the PL spectra of QD-MoS$_2$ on SiO$_2$  at 20 K at high laser powers about, we did not observe any multiple peaks. Fig.~\ref{fig:4}(d) shows the PL spectra of QD-MoS$_2$ on hBN  at the same T and same power, we can clearly observe multiple PL emissions which are resolved by fitting the PL spectra using multiple Lorentzian.

       As we discussed earlier, there is a possibility of the formation of   interlayer exciton in this structure. We try to understand the multiple emission mechanism in terms of the interlayer excitons.
       There is a possibility that these interesting peaks may be the signature of Moor{\'e} excitons which are the interlayer excitons formed in the MoS$_2$ and QD structure trapped   in the possibly formed Moor{\'e} potential due to the crystal plane mismatch between MoS$_2$ and h-BN. However, these interesting observations need to be further explored in detail from Moor{\'e} potential and other aspects.

    It has been shown that if hBN is used as a capping layer in MoSe$_2$/WSe$_2$ heterostructure, the inhomogeneous PL linewidths will reduce giving rise to equal energy  spaced interlayer excitons at lower temperatures which can be attributed to Moor{\'e} excitons.\cite{tran2019evidence} It is also interesting to note that these kind of inhomogeneous broadening in the presence of hBN may also occur due the presence of multiple Moor{\'e} domains or strain caused by the substrate and the interlayer spacing within the laser spot. In addition, the multiple excitons peaks in the PL spectra can also be observed in the heterostructure due the quantized energy levels caused by the confinement effects.   \cite{tran2019evidence,torchynska2005ground} 
    
       The observation of multiple PL peaks  depends on various factors. Firstly, as discussed previously, the amount of lattice mismatch between MoS$_2$ and hBN and whether this mismatch can create Moor{\'e} potentials. secondly the strain created by this match and its effect on the PL spectra.\cite{marzin1994calculation,cusack1997absorption} This strain field which is mostly biaxial in nature and is effective on the entire area of the QD covered on the MoS$_2$-hBN which can lead to multiple confined electronic level.\cite{thoai1990influence,grundmann1995inas} This strain effect can also be the reason for the observed multiple peaks in the spectra.
    These kind of multiple PL emission of QD combined with quantum well which are similar to 2D semiconductors in many ways are observed at low T at higher laser powers.\cite{torchynska2005ground} However further studies is expected in these directions.

  In summary, we measured the PL spectra of MoS$_2$-QD heterostructure on SiO$_2$ and on hBN at low temperatures. At low T, we observe multiple PL emission peaks of the heterostructure on hBN which are absent on SiO$_2$. These multiple peaks may be arising due to the lattice match between MoS$_2$ and hBN or the strain field created by the MoS$_2$-hBN or due the altered quantum potential of QD due the presence of hBN. Further explanation and detailed lifetime and other studies are expected in near future.   

\textbf{Acknowledgment:} Author thanks CSIR-UGC for financial support and  DST Nanomission for funding. Author thanks Aveek Bid and Jaydeep Kumar Basu for discussion. Author thanks Komal Sharma for helping in QD synthesis. Author thanks the   facilities of CeNSE and IISc.

* This manuscript was the residue results of the Ph.D work of the author under the guidance of Aveek Bid and Jaydeep Kumar Basu in IISc.

	  

\bibliography{ref}

\end{document}